\documentclass[runningheads]{llncs}
\usepackage[T1]{fontenc}
\usepackage{graphicx}
\usepackage{cleveref}
\usepackage{booktabs}
\usepackage{array}
\usepackage{tabularx}

\newcommand\blfootnote[1]{%
  \begingroup
  \renewcommand\thefootnote{}\footnote{#1}%
  \addtocounter{footnote}{-1}%
  \endgroup
}

\begin{document}
\title{Large-Scale Knowledge Integration for Enhanced Molecular Property Prediction}
\titlerunning{Large-Scale Knowledge Integration}
%
\author{Yasir Ghunaim\inst{1}$^{*}$\orcidID{0000-0003-4955-3813} \and 
Robert Hoehndorf\inst{1}\orcidID{0000-0001-8149-5890}} 
\authorrunning{Y. Ghunaim et al.}
%
\institute{$^1$ King Abdullah University of Science and Technology, Thuwal, Saudi Arabia\\
\email{\{yasir.ghunaim,robert.hoehndorf\}@kaust.edu.sa}}
\maketitle              
\begin{abstract}
Pre-training machine learning models on molecular properties has proven effective for generating robust and generalizable representations, which is critical for advancements in drug discovery and materials science. While recent work has primarily focused on data-driven approaches, the KANO model introduces a novel paradigm by incorporating knowledge-enhanced pre-training. In this work, we expand upon KANO by integrating the large-scale ChEBI knowledge graph, which includes 2,840 functional groups -- significantly more than the original 82 used in KANO. We explore two approaches, Replace and Integrate, to incorporate this extensive knowledge into the KANO framework. Our results demonstrate that including ChEBI leads to improved performance on 9 out of 14 molecular property prediction datasets. This highlights the importance of utilizing a larger and more diverse set of functional groups to enhance molecular representations for property predictions. \blfootnote{$^{*}$Corresponding author.}
\blfootnote{Code: github.com/Yasir-Ghunaim/KANO-ChEBI}

\keywords{Knowledge-enhanced Learning \and Knowledge Graph  \and Molecular Property Prediction.}
\end{abstract}
\section{Introduction}
Large-scale pre-trained machine learning models have become the gold standard in fields such as computer vision and natural language processing~\cite{brock2018large,devlin2018bert,brown2020language,radford2021learning}, demonstrating remarkable performance across a diverse array of tasks. For instance, models like DALL-E~\cite{ramesh2021zero} can generate detailed images in seconds, tasks that would otherwise take expert graphic designers hours or days to accomplish. Similarly, language models like ChatGPT~\cite{brown2020language} can resolve complex programming challenges, engage in human-like conversations, and distill knowledge from extensive texts. Given these impressive capabilities, there is growing interest in leveraging such models in scientific domains, including chemistry and biology.

In the field of molecular property prediction, continuous improvements have been made using representations derived from available datasets. However, much of the existing work has focused solely on data-driven approaches in the scientific domain. An emerging paradigm, exemplified by the KANO~\cite{fang2023knowledge} model, seeks to enhance these representations by integrating well-established scientific knowledge into the pre-training process. KANO creates a knowledge graph, called ElementKG, of atomic data sourced from the periodic table and a small collection of 82 functional groups extracted from Wikipedia. This integration is crucial since functional groups play a pivotal role in chemistry as they dictate molecular behavior regarding chemical properties such as toxicity or reactivity~\cite{ertl2020most}. For instance, compounds containing the Cyano group are often toxic~\cite{kim2012synthesis}, which highlights the importance of identifying such functional groups to improve molecular property predictions. By guiding both the pre-training and fine-tuning processes, ElementKG enables KANO to achieve state-of-the-art results across 14 molecular property prediction tasks.

However, KANO's reliance on a relatively narrow set of functional groups limits its ability to capture the full diversity of chemical structures. In contrast, comprehensive knowledge graphs like ChEBI~\cite{degtyarenko2007chebi} and PubChem~\cite{kim2016pubchem} have much broader collections, with ChEBI containing 2,840 functional groups -- nearly 35 times more than ElementKG. We hypothesize that integrating this larger and more diverse set of functional groups could further improve the chemical diversity and performance of knowledge-enhanced molecular representations.

In this paper, we extend the KANO framework to incorporate the extensive functional group knowledge from the ChEBI knowledge graph.
We propose two approaches --- Replace and Integrate --- to incorporate this knowledge into the molecular representations learned by KANO. Through experiments on 14 benchmark datasets, we find that leveraging a broader collection of functional groups can lead to improvements in predictive performance on certain datasets. These results suggest the potential benefits of integrating large-scale knowledge graphs for molecular property prediction and motivate further research in knowledge-enhanced learning for scientific applications. Our work takes a step towards more effective utilization of existing scientific knowledge to improve molecular property prediction.

\section{Related work}
In this review, we classify the literature on pre-training for molecular property prediction into two main categories: data-driven and knowledge-enhanced pre-training.

\textbf{Data-driven Pre-training.} Data-driven approaches have predominantly shaped the pre-training landscape for molecular data~\cite{liu2019n,hu2019strategies,rong2020self,zhang2021motif,liu2021pre,fang2022geometry,wang2022molecular}. Hu et al.\cite{hu2019strategies} introduced influential strategies, proposing both node and graph-level tasks to learn discriminative representations. While their node-level task employed a self-supervised method, the graph-level task was supervised, which limited pre-training scalability to the availability of labeled data. Building upon this, GROVER\cite{rong2020self} incorporated an edge-level task and a self-supervised motif prediction approach, enabling pre-training on larger unlabeled datasets. Subsequent works explored various molecular features and pre-training frameworks, such as motif topology~\cite{zhang2021motif}, geometry~\cite{liu2021pre,fang2022geometry}, and contrastive learning~\cite{wang2022molecular}.

\begin{figure}[t!]
    \centering
    \includegraphics[width=\textwidth]{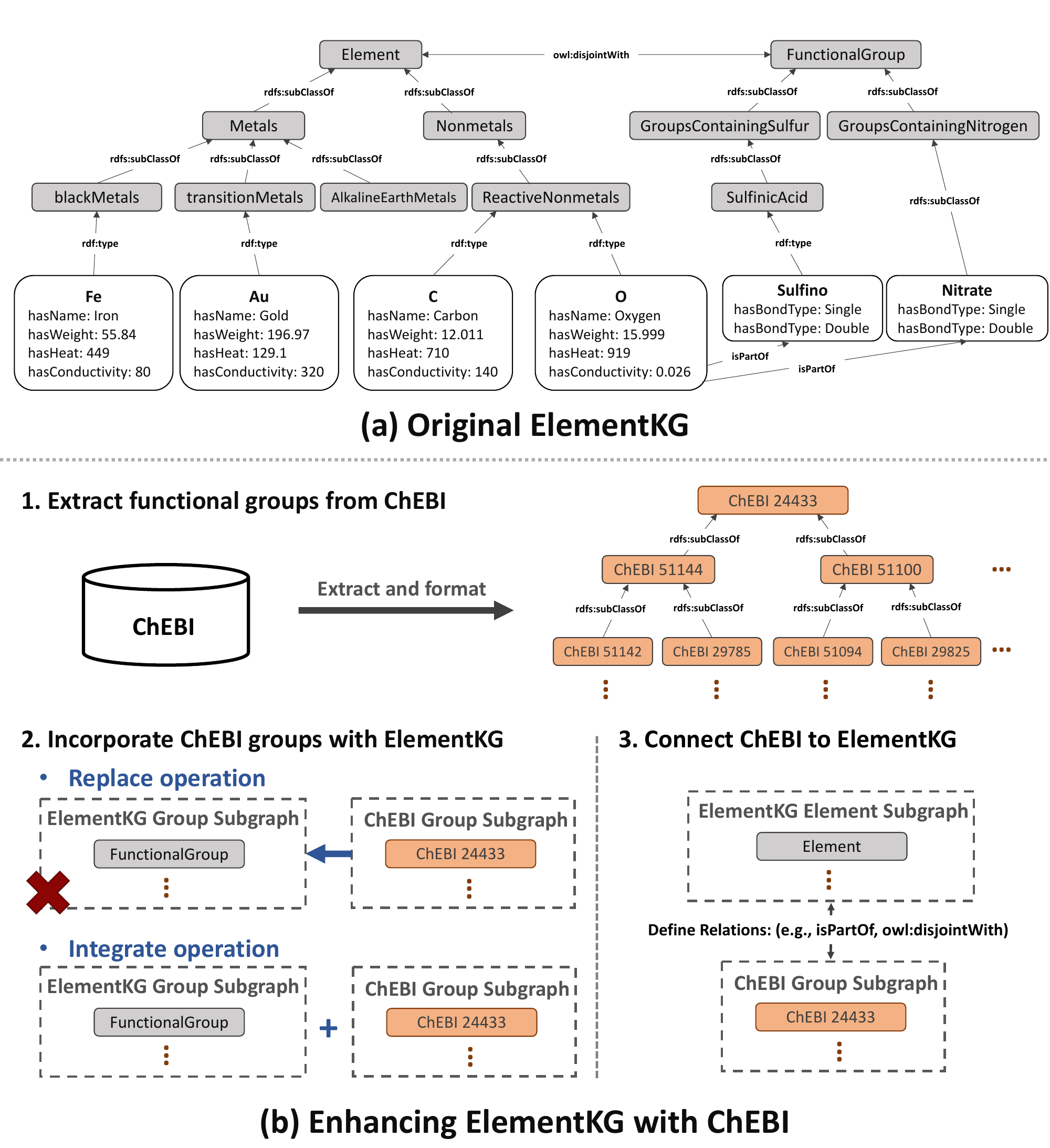}
    \caption{\textbf{(a) Original ElementKG structure utilized by
        KANO.} Elements and functional groups derived from the periodic table and Wikipedia. 
      \textbf{(b) Enhancing ElementKG with ChEBI.}
      Our methodology involves extracting functional groups from ChEBI and incorporating them into ElementKG using Replace or Integrate operations. The Replace operation removes the \texttt{FunctionalGroup} subgraph and replaces it with ChEBI groups, while the Integrate operation adds new ChEBI groups without removing existing data. Finally, we define relations between ChEBI groups and entities in the \texttt{Element} subgraph of ElementKG.}
    \label{fig:setup}
\end{figure}

\textbf{Knowledge-enhanced Pre-training.} 
While data-driven approaches have achieved impressive results, their generalizability can be limited by the specific datasets used. In contrast, knowledge-enhanced pre-training leverages scientific information from knowledge graphs to improve generalization. KANO~\cite{fang2023knowledge}, which established state-of-the-art results across 14 datasets, represented significant progress in this area. However, KANO used a comparatively small knowledge graph of molecular functional groups. Our work extends upon KANO by leveraging large-scale knowledge graphs, such as ChEBI~\cite{degtyarenko2007chebi}, to enhance the chemical diversity and scale of functional groups used in pre-training, thereby aiming to improve predictive performance.

\section{Methodology}

\subsection{KANO Pre-training and Fine-tuning using ElementKG}
\label{kano_methodology}
In this section, we provide a brief overview of KANO's pre-training and fine-tuning processes utilizing ElementKG. For a detailed explanation, please refer to the original KANO paper~\cite{fang2023knowledge}. 

\sloppy As previously described, KANO is initially set up by constructing ElementKG, a knowledge graph composed of two main subgraphs: \texttt{Element} and \texttt{FunctionalGroup}, as illustrated in \Cref{fig:setup}(a). The \texttt{Element} subgraph contains atomic data derived from the periodic table, while the \texttt{FunctionalGroup} subgraph includes information on functional groups sourced from Wikipedia. Following the creation of ElementKG, KANO generates embeddings from the knowledge graph, which are utilized to enhance molecular representations during both pre-training and fine-tuning stages.

KANO employs a self-supervised learning approach during pre-training, utilizing contrastive learning. This method involves generating an augmented version of each input molecular graph by incorporating atomic-level chemistry knowledge from the \texttt{Element} subgraph of ElementKG. The model is then trained to maximize agreement between the original and augmented graphs while minimizing similarity with other graphs in the training minibatch. This pre-training process creates knowledge-enhanced representations that are used to improve the fine-tuning on prediction tasks.

During the fine-tuning phase for molecular property prediction tasks, the process starts with the functional group detection step, which identifies functional groups within each given input molecular graph. KANO then retrieves the corresponding ElementKG embeddings of the detected functional groups and combines these embeddings into a single prompt. This prompt is integrated into the atom representations of the molecular graph to enhance the input representations. The enriched graph is then used in the fine-tuning process for the target property prediction tasks.

\subsection{Enhancing ElementKG with ChEBI}
\label{proposed_methodology}
The original ElementKG knowledge graph, utilized by KANO, shown in \Cref{fig:setup}(a), includes atomic information from the periodic table and 82 functional groups sourced from Wikipedia. While ElementKG provides comprehensive atomic data, its diversity of functional groups is somewhat limited, especially when compared to the 2,840 functional groups in ChEBI. To improve the chemical diversity of the KANO model's representations, we extend ElementKG to integrate the broader set of functional groups from ChEBI.

\begin{figure}[t]
    \centering
    \includegraphics[width=\textwidth]{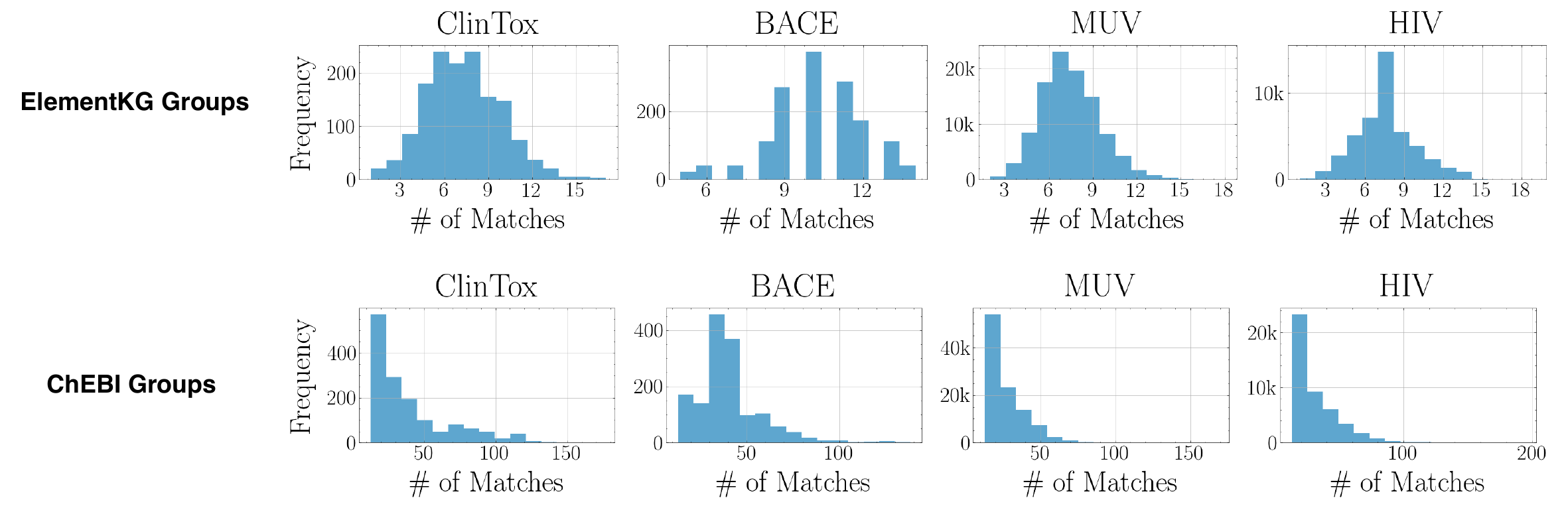}
    \caption{\textbf{Histogram comparing functional group matches for
        KANO using ElementKG versus ChEBI.} The histograms display the
      distribution of functional group matches in different datasets
      (ClinTox, BACE, MUV, HIV) when using ElementKG and
      ChEBI. Notably, ChEBI's larger functional group set results in a
      significantly higher number of matches across all datasets,
      illustrating its potential for more detailed molecular
      characterization.}
    \label{fig:histogram}
\end{figure}

We first extract the functional group subgraph from ChEBI and reformat it to align with ElementKG's structure. To combine the ChEBI subgraph with ElementKG, we develop two main operations: Replace and Integrate, as highlighted in \Cref{fig:setup}(b). 
The Replace operation removes the existing \texttt{FunctionalGroup} subgraph from ElementKG and its associated relations before adding the ChEBI functional group subgraph. In contrast, the Integrate operation incorporates the ChEBI subgraph without removing any existing data. For both operations, we establish relations (i.e., \texttt{isPartOf}) between the newly added functional group subgraph and the existing \texttt{Element} entities in ElementKG. Specifically, we extract unique atom types from the SMILES strings provided in ChEBI groups and use these to determine which \texttt{Element} entities to create new relations with. Following KANO, we also define axioms (e.g., \texttt{ owl:disjointWith}) to distinguish between different classes of \texttt{Element} and ChEBI groups.

After expanding ElementKG to include ChEBI, we revisited KANO’s fine-tuning process to accommodate the increased diversity of functional groups. As described in Section \ref{kano_methodology}, during fine-tuning, KANO performs a functional group detection step to identify the groups present in each molecule input. The original setup restricts functional group matches to 13 per molecule input. However, as shown in \Cref{fig:histogram}, most molecules typically exhibit fewer than 13 matches when using ElementKG groups alone, with a few exceptions. With the integration of ChEBI groups, we observe a significant increase in matches, with some inputs having more than 100. To fully leverage this chemical information, we remove the 13-match limit, ensuring all detected functional groups are used during the model's fine-tuning process.

\section{Experiments}

\textbf{Baselines:} As we are building upon KANO~\cite{fang2023knowledge}, our main comparison is against its original implementation. We refer to the original implementation using ElementKG functional groups as ``KANO, E'', where `E' stands for ElementKG. To ensure consistency across experiments, we re-ran ``KANO, E'' using the same hyperparameters as the original implementation of KANO. Note that our runs have slightly different values than the ones reported in the original paper.  In addition, we also evaluate a variant of KANO that omits the 13-match limit during functional group detection, referred to as ``KANO*, E''. For the ``Replace'' and ``Integrate'' operations described in Section \ref{proposed_methodology}, we establish two baselines:
\begin{itemize}
\item ``KANO*, C'': The ``Replace'' baseline, which utilizes ChEBI functional groups exclusively (`C' stands for ChEBI).
\item ``KANO*, E+C'': The ``Integrate'' baseline, combining functional groups from both ElementKG and ChEBI.
\end{itemize}

\noindent\textbf{Tasks and evaluation:} Our evaluation framework utilizes the same tasks and metrics as KANO, covering 14 datasets from the MoleculeNet~\cite{wu2018moleculenet} benchmark. These datasets are divided into classification and regression tasks:
\begin{itemize}
\item The classification tasks are BBBP, Tox21, ToxCast, SIDER, ClinTox, BACE, MUV, and HIV. These tasks span biophysics and physiology domains. Following KANO, we evaluate these tasks using the receiver operating characteristic - area under curve (ROC-AUC\%) metric.
\item The regression tasks are ESOL, FreeSolv, Lipophilicity, QM7, QM8 and QM9, which focus on physical chemistry and quantum mechanics. Similar to KANO, we use the root-mean-square error (RMSE) metric for ESOL, FreeSolv and Lipophilicity, while we use the mean absolute error (MAE) for QM7, QM8 and QM9.
\end{itemize}

\noindent\textbf{Results:} 
The results are divided into two categories based on the dataset type: classification and regression. The classification dataset results are presented in \Cref{tab:performance1}, while the regression dataset results are reported in \Cref{tab:performance2}.

In the classification benchmarks, KANO variants that use ChEBI functional groups improve performance in 6 out of 8 datasets, showcasing the benefits of expanding the size and diversity of functional groups. However, in the ClinTox dataset, using only ChEBI functional groups leads to a performance degradation, suggesting that ClinTox may specifically benefit from the unique functional group types in ElementKG. For the HIV dataset, both ChEBI variants outperform the original KANO (KANO, E), but KANO* with the ElementKG configuration (KANO*, E) achieves the highest performance. This suggests that removing the 13-match limit may offer a particular advantage for some datasets, such as HIV and ClinTox.

\begin{table}[t]
\caption{\textbf{Effect of ChEBI Integration on Classification Tasks.} The mean of the ROC-AUC (\%) averaged over three runs (higher is better).}
\centering
\fontsize{7pt}{10pt}\selectfont
\begin{tabular}{@{}p{1.8cm}*{8}{>{\centering\arraybackslash}p{1.2cm}}@{}}
    \toprule
     & BBBP & Tox21 & ToxCast & SIDER & ClinTox & BACE & MUV & HIV \\ 
    \midrule 
    KANO, E & 94.1$_{\pm 0.8}$ & 82.1$_{\pm 2.4}$ & 70.2$_{\pm 1.0}$ & 61.9$_{\pm 2.8}$ & 91.0$_{\pm 4.9}$ & 87.3$_{\pm 4.2}$ & 80.5$_{\pm 2.0}$ & 80.7$_{\pm 1.8}$ \\
    KANO*, E & 93.5$_{\pm 0.9}$ & 82.0$_{\pm 2.1}$ & 69.8$_{\pm 1.6}$ & 61.2$_{\pm 2.2}$ & \textbf{91.2$_{\pm 3.4}$} & 87.0$_{\pm 4.6}$ & 79.6$_{\pm 1.7}$ & \textbf{82.6$_{\pm 1.8}$} \\
    KANO*, C & 93.8$_{\pm 0.9}$ & \textbf{82.6$_{\pm 2.6}$} & \textbf{70.5$_{\pm 1.3}$} & \textbf{62.1$_{\pm 1.5}$} & 90.3$_{\pm 4.2}$ & \textbf{88.3$_{\pm 3.2}$} & 81.1$_{\pm 1.4}$ & 81.4$_{\pm 2.8}$ \\
    KANO*, E+C & \textbf{94.6$_{\pm 1.3}$} & 82.2$_{\pm 3.7}$ & 70.2$_{\pm 1.0}$ & 59.9$_{\pm 2.6}$ & 91.0$_{\pm 4.4}$ & 87.7$_{\pm 4.5}$ & \textbf{81.8$_{\pm 0.7}$} & 81.7$_{\pm 3.1}$ \\ 
    \midrule
\end{tabular}
\label{tab:performance1}
\end{table}

\begin{table}[t]
\caption{\textbf{Effect of ChEBI Integration on Regression Tasks.} Mean RMSE for ESOL, FreeSolv, Lipophilicity and MAE for QM7, QM8, QM9, averaged over three runs (lower is better).}
\centering
\fontsize{7pt}{10pt}\selectfont
\begin{tabular}{@{}p{1.8cm}*{4}{>{\centering\arraybackslash}p{1.5cm}}*{1}{>{\centering\arraybackslash}p{1.7cm}}*{1}{>{\centering\arraybackslash}p{1.95cm}}@{}}
    \toprule
     & ESOL & FreeSolv & Lipophilicity & QM7 & QM8 & QM9 \\ 
    \midrule
    KANO, E & 0.776$_{\pm 0.089}$ & 1.593$_{\pm 0.633}$ & 0.610$_{\pm 0.025}$ & 64.31$_{\pm 6.48}$ & 0.0153$_{\pm 0.001}$ & 0.00433$_{\pm 0.0001}$ \\
    KANO*, E & 0.766$_{\pm 0.096}$ & 1.595$_{\pm 0.634}$ & \textbf{0.600$_{\pm 0.022}$} & \textbf{63.33$_{\pm 6.98}$} & 0.0154$_{\pm 0.001}$ & \textbf{0.00430$_{\pm 0.0001}$} \\
    KANO*, C & 0.747$_{\pm 0.068}$ & 1.675$_{\pm 0.684}$ & 0.612$_{\pm 0.022}$ & 64.28$_{\pm 7.12}$ & \textbf{0.0150$_{\pm 0.001}$} & 0.00437$_{\pm 0.0001}$ \\
    KANO*, E+C & \textbf{0.740$_{\pm 0.058}$} & \textbf{1.584$_{\pm 0.678}$} & 0.601$_{\pm 0.019}$ & 64.23$_{\pm 5.43}$ & 0.0153$_{\pm 0.001}$ & 0.00436$_{\pm 0.0001}$ \\
    
    \midrule
\end{tabular}
\label{tab:performance2}
\end{table}

In the regression benchmarks, the integration of the larger ChEBI functional group set showed slight improvements in the ESOL, FreeSolv, and QM8 datasets, outperforming the ElementKG-only version. However, results in the Lipophilicity, QM7, and QM9 datasets did not show any improvements. This variance in performance could be attributed to the adequacy of the original ElementKG functional groups, which may already cover the necessary molecular features for these tasks. Alternatively, the current model architecture might not be fully optimized to utilize the expanded chemical knowledge provided by ChEBI. These outcomes indicate a task-specific dependency, where the benefits of integrating ChEBI's functional groups depend significantly on the characteristics of each dataset and the specific requirements of the prediction tasks.

\noindent\textbf{Limitations:} Although our results demonstrate the benefits of the proposed modification operations, Replace and Integrate, we observe that their effectiveness is sensitive to both task type (classification vs. regression) and specific datasets. Further research is necessary to understand these sensitivities and optimize our methods accordingly. One potential area for improvement is the handling of overlapping functional groups between ElementKG and ChEBI. Developing a more refined integration mechanism to effectively merge such overlaps could lead to better knowledge representation and, consequently, improved model performance. Additionally, while we focus on integrating ChEBI in this work, exploring the incorporation of other large-scale knowledge graphs, such as PubChem, might yield further performance enhancements by providing complementary chemical information. Addressing these limitations and expanding the scope of knowledge integration are promising directions for future research in knowledge-enhanced molecular property prediction.

\section{Conclusion}
In this work, we integrated the large-scale ChEBI knowledge graph into the KANO framework for molecular property prediction. Our Replace and Integrate approaches show that expanding the functional group diversity can improve model performance on various classification and regression tasks, highlighting the potential of knowledge-enhanced learning in molecular science. However, our results also reveal sensitivities to specific task types and datasets, emphasizing the need for further research. Despite these challenges, our work demonstrates the promise of knowledge-enhanced learning in molecular property prediction and lays the foundation for future advancements in drug design, materials science, and other applications.

\begin{credits}
\subsubsection{\ackname} This work has been supported by funding from King Abdullah University of Science and Technology (KAUST) Office of Sponsored Research (OSR) under Award No. URF/1/4675-01-01, URF/1/4697-01-01, URF/1/5041-01-01, REI/1/5659-01-01, REI/1/5235-01-01, and FCC/1/1976-46-01.  Yasir Ghunaim is funded by Saudi Aramco. We acknowledge support from the KAUST Supercomputing Laboratory.

\end{credits}

%
%
%
%
\bibliographystyle{splncs04}
\bibliography{mybibliography}

\end{document}